\documentclass[sigconf]{acmart}

\copyrightyear{2025}
\acmYear{2025}
\setcopyright{acmlicensed}\acmConference[CompEd 2025]{Proceedings of the ACM 
Global Computing Education Conference 2025 Vol 1}{October 21--25, 
2025}{Gaborone, Botswana}
\acmBooktitle{Proceedings of the ACM Global Computing Education Conference 2025 
Vol 1 (CompEd 2025), October 21--25, 2025, Gaborone, Botswana}
\acmDOI{10.1145/3736181.3747131}
\acmISBN{979-8-4007-1929-5/2025/10}

\usepackage{xspace}
\usepackage[capitalise]{cleveref}
\usepackage{hyperref}
\usepackage{subcaption}
\usepackage[
    colorinlistoftodos,prependcaption,textsize=tiny
]{todonotes}
\usepackage{makecell}

\usepackage{scratch3}
\usepackage{tabularx}
\usepackage{colortbl}
\usepackage{array}

\newcommand{\parag}[1]{\vspace{0.08em}\noindent {\bf #1}}

\newcommand{\summary}[2]{
	\vspace{0.4em}
	\noindent
	\colorbox{gray!20}{%
		\parbox{.97\linewidth}{%
			\textbf{\textsf{Summary (\textit{#1})}}
			#2
		}%
	}%
}%

\usepackage{tikz}
\usepackage{pgflibraryshapes}
\usetikzlibrary{arrows}
\usetikzlibrary{chains}
\usetikzlibrary{fadings}
\usetikzlibrary{matrix}
\usetikzlibrary{fit}
\usetikzlibrary{positioning}
\usetikzlibrary{shapes}
\usetikzlibrary{automata}
\tikzset{
	ctrlstate/.style = {state,align=center,inner sep=2pt, minimum size=2mm},
	cfastate/.style = {ctrlstate},
	cfatargetstate/.style = {cfastate, double},
	compstate/.style = {cfastate, rounded rectangle, minimum height=5mm, 
	minimum width=3em, inner sep=3pt},
	concept/.style = {cfastate, inner sep=3pt, fill=gray!50, rectangle,
		minimum width=17mm, minimum height=9mm, draw=gray!6 },
	crosscutting/.style = {concept, rounded rectangle, fill=gray!30},
	conceptstate/.style = {concept, rounded rectangle, fill=gray!30, 
	draw=gray!90, minimum width=20mm},
	inputstate/.style = {concept, rectangle, fill=gray!10, draw=gray!90, 
	minimum width=20mm, inner sep=3pt},
	explain/.style = {circle, draw=gray!30, line width=1mm, minimum size=7mm},
	one/.style = {fill=blue!70,draw=blue!70},
	two/.style = {fill=red!50,draw=red!50},
	abststate/.style = {rectangle,align=center,inner sep=2pt,minimum 
	size=3.5mm,fill=gray!0, draw=gray!90},
	line/.style = {draw},
	trans/.style = {draw,semithick,->,shorten >=1pt,>=stealth'},
	missing/.style = {draw=red,densely dotted,fill=red,semithick,->,shorten 
	>=1pt,>=stealth'},
	ctrans/.style = {draw,very thick,->,shorten >=1pt,>=stealth',draw=gray!90},
	epsilon/.style = {trans,dashed},
	strengthen/.style = {draw=gray!30,semithick,double,shorten 
	>=1pt,>=stealth',line width=1mm},
}

\newcommand*\circled[1]{\tikz[baseline=(char.base)]{
    \node[shape=circle,draw,inner sep=0.4pt] (char) {\textsf{\small#1}};}}

\newcommand{\motionblockleft}{\begin{mbox}\sf\begin{tikz}[baseline=(X.base)]\node[draw=black!60,fill=blue!12,semithick,rectangle,inner
 sep=1pt, minimum size=1em, outer sep=0pt, rounded corners=1pt] (X)}%
\newcommand{\controlblockleft}{\begin{mbox}\sf\begin{tikz}[baseline=(X.base)]\node[draw=black!60,fill=orange!15,semithick,rectangle,inner
 sep=1pt, minimum size=1em, outer sep=0pt, rounded corners=1pt] (X)}%
\newcommand{\hatblockleft}{\begin{mbox}\sf\begin{tikz}[baseline=(X.base)]\node[draw=black!60,fill=yellow!20,semithick,rectangle,inner
 sep=1pt, minimum size=1em, outer sep=0pt, rounded corners=1pt] (X)}%
\newcommand{\looksblockleft}{\begin{mbox}\sf\begin{tikz}[baseline=(X.base)]\node[draw=black!60,fill=violet!20,semithick,rectangle,inner
 sep=1pt, minimum size=1em, outer sep=0pt, rounded corners=1pt] (X)}%
\newcommand{\sensingblockleft}{\begin{mbox}\sf\begin{tikz}[baseline=(X.base)]\node[draw=black!60,fill=cyan!20,semithick,rectangle,inner
 sep=1pt, minimum size=1em, outer sep=0pt, rounded corners=1pt] (X)}%
\newcommand{\soundblockleft}{\begin{mbox}\sf\begin{tikz}[baseline=(X.base)]\node[draw=black!60,fill=magenta!20,semithick,rectangle,inner
 sep=1pt, minimum size=1em, outer sep=0pt, rounded corners=1pt] (X)}%
\newcommand{\operatorblockleft}{\begin{mbox}\sf\begin{tikz}[baseline=(X.base)]\node[draw=black!60,fill=green!20,semithick,rectangle,inner
 sep=1pt, minimum size=1em, outer sep=0pt, rounded corners=1pt] (X)}%

\newcommand{\blockleft}{\begin{mbox}\sf\begin{tikz}[baseline=(X.base)]\node[draw=black!60,fill=black!3,semithick,rectangle,inner
 sep=1pt, minimum size=1em, outer sep=0pt, rounded corners=1pt] (X)}%
\newcommand{\blockright}{;\end{tikz}\normalfont\end{mbox}}%

\newcommand{\mblockleft}{\begin{mbox}\sf\begin{tikz}[baseline=(X.base)]\node[draw=red!60,densely
 dotted,fill=red!3,semithick,rectangle,inner sep=1pt, minimum size=1em, outer 
sep=0pt, rounded corners=1pt] (X)}%
\newcommand{\mblockright}{;\end{tikz}\normalfont\end{mbox}}%

\newcommand\definetool[2]{\newcommand{#1}{{\textsc{#2}}\xspace}}
\definetool{\Scratch}{Scratch}
\definetool{\litterbox}{LitterBox}
\definetool{\drscratch}{Dr. Scratch}
\definetool{\qualityhound}{Qualityhound}
\definetool{\hairball}{Hairball}
\definetool{\itch}{Itch}
\definetool{\whisker}{Whisker}
\definetool{\jask}{Jask}

\newcommand{\inlineFigure}[2]{$\vcenter{\hbox{\includegraphics[scale=#1]{#2}}}$}%
\newcommand{\numbertype}{\inlineFigure{0.025}{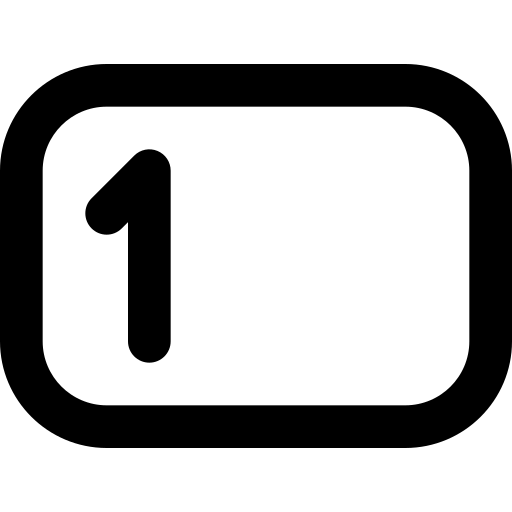}\xspace}
\newcommand{\stringtype}{\inlineFigure{0.02}{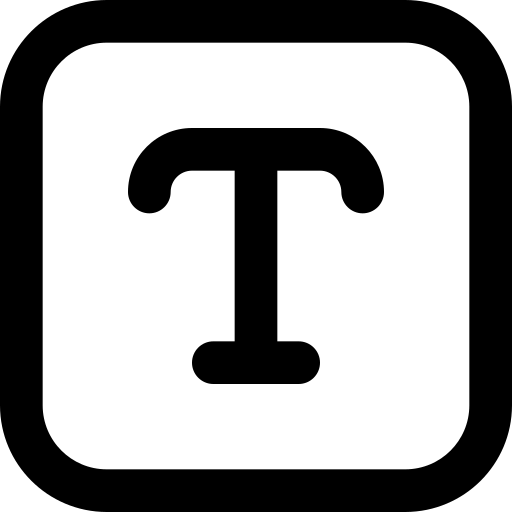}\xspace}
\newcommand{\yesnotype}{\inlineFigure{0.02}{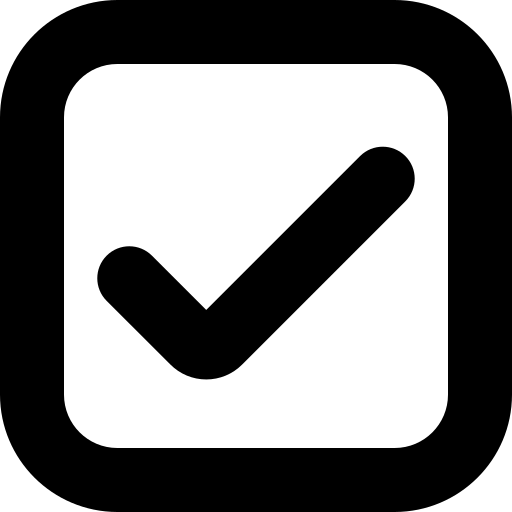}\xspace}
\newcommand{\multipletype}{\inlineFigure{0.55}{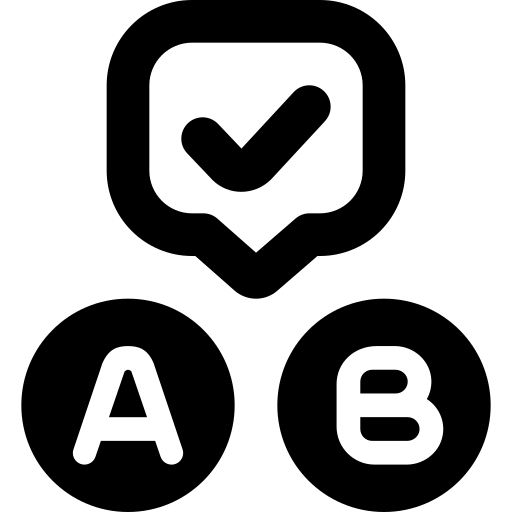}\xspace}
\newcommand{\freetype}{\inlineFigure{0.55}{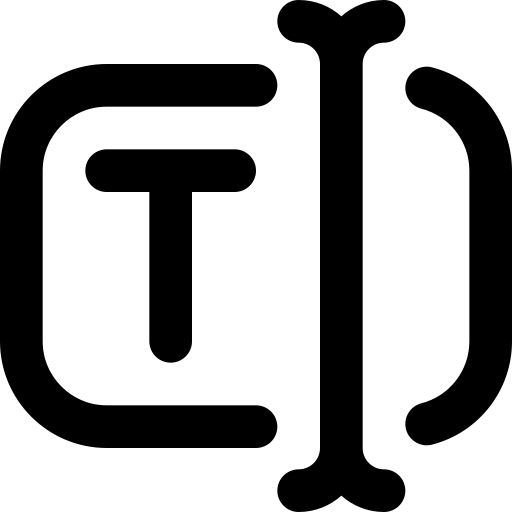}\xspace}

\newcommand{\numprojects}{600,913\xspace}
\newcommand{\numquestions}{54,118,694\xspace}

\newcommand{\atomTextProject}{64.5\%\xspace}
\newcommand{\atomExecutionProject}{30.4\%\xspace}
\newcommand{\atomPurposeProject}{49.0\%\xspace}
\newcommand{\blockTextProject}{61.2\%\xspace}
\newcommand{\blockExecutionProject}{58.4\%\xspace}
\newcommand{\blockPurposeProject}{100.0\%\xspace}
\newcommand{\relationTextProject}{46.5\%\xspace}
\newcommand{\relationExecutionProject}{99.0\%\xspace}
\newcommand{\relationPurposeProject}{31.2\%\xspace}
\newcommand{\macroTextProject}{100.0\%\xspace}
\newcommand{\macroExecutionProject}{71.1\%\xspace}
\newcommand{\macroPurposeProject}{100.0\%\xspace}

\begin{document}
\title{Automatically Generating Questions About Scratch Programs}
\settopmatter{authorsperrow=2}
\author{Florian Obermüller}
\email{obermuel@fim.uni-passau.de}
\orcid{0000-0002-6752-6205}
\affiliation{%
  \institution{University of Passau}
  \country{Germany}
}

\author{Gordon Fraser}
\email{gordon.fraser@uni-passau.de}
\orcid{0000-0002-4364-6595}
\affiliation{%
  \institution{University of Passau}
  \country{Germany}
}


\begin{abstract}
  When learning to program, students are usually assessed based on the
  code they wrote. However, the mere completion of a programming task
  does not guarantee actual comprehension of the underlying
  concepts. Asking learners questions about the code they wrote has
  therefore been proposed as a means to assess program
  comprehension. As creating targeted questions for individual student
  programs can be tedious and challenging, prior work has proposed to
  generate such questions automatically.
  In this paper we generalize this idea to the block-based programming
  language \Scratch. We propose a set of 30 different questions for
  \Scratch code covering an established program comprehension model,
  and extend the \litterbox static analysis tool to automatically
  generate corresponding questions for a given \Scratch program.
  %
On a dataset of \numprojects projects we generated 
\numquestions questions automatically.
  Our initial experiments with 34 ninth graders
  demonstrate that this approach can indeed generate meaningful
  questions for \Scratch programs, and we find that the ability of
  students to answer these questions on their programs relates to
  their overall performance.
  \end{abstract}




\maketitle
\section{Introduction}
\label{sec:introduction}

\begin{figure}[t]
  \centering
  \subfloat[\label{fig:code}\centering Student code
  ]{{\includegraphics[width=.48\columnwidth]{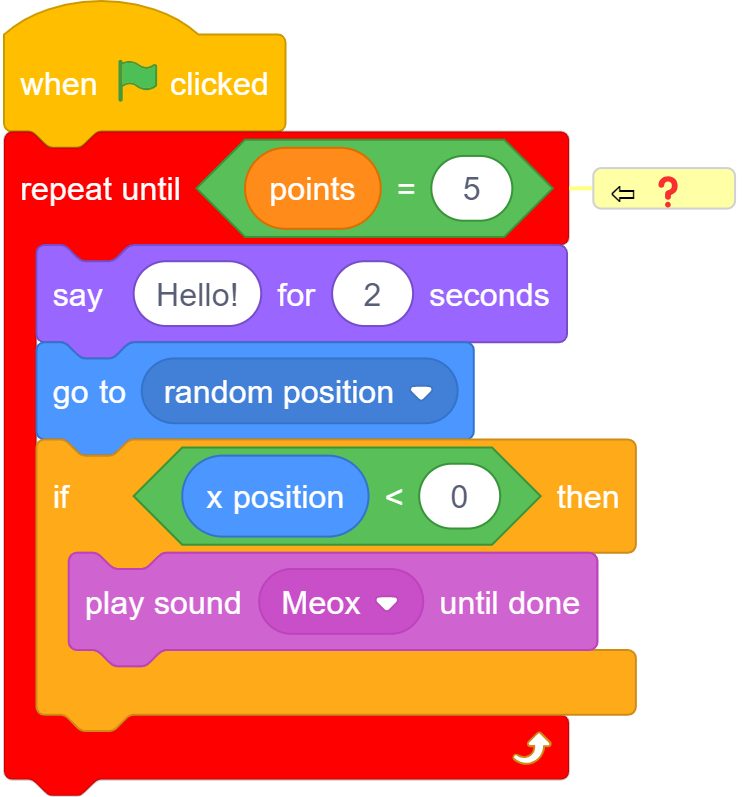}}}
  \hfill
  \subfloat[\label{fig:question}\centering  Answer choices
  ]{{\includegraphics[width=.3\columnwidth]{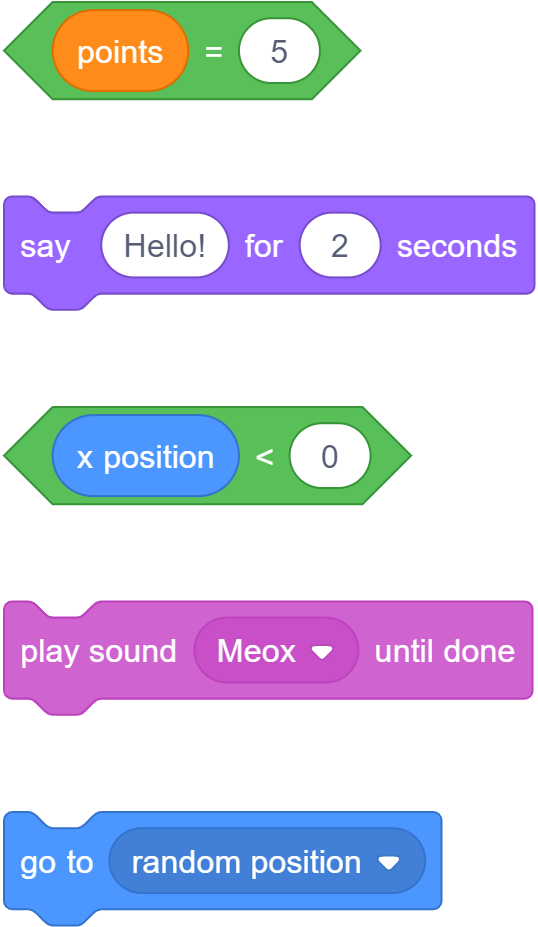}}}
  %
  \caption{\label{fig:question_example} Example for the question 
    \textit{Which of these blocks controls how many times this loop is 
      executed?}}
\end{figure}

A primary means of assessing student learning in programming education
is through the code they write. While the ability to write code is an
essential skill, it does not necessarily correspond to an
understanding of the programming concepts being
used~\cite{lopez2008relationship}. Indeed the completion of a
programming task does not even guarantee that the student actually
understands what their own program does~\cite{lehtinen2021struggle}.
In order to better assess program comprehension skills, one proposed
solution is to ask students questions about the code they wrote.  Such
\emph{Questions about Learner’s Code (QLCs)} have proven to be
suitable indicators of understanding and a better metric than simply
checking if the final program solution is
correct~\cite{lehtinen2021struggle}.

One challenge in using QLCs is that questions need to be derived
individually for each student's solution. This can be a tedious and
time-consuming task given large cohorts of learners, and neither
scales to in-class feedback, nor to post-submission assessment. To
address this issue, it has been suggested to automatically generate
QLCs from students’ code~\cite{lehtinen2021askAutomatically}. This has
been investigated in the context of textual programming
languages~\cite{lehtinen2023questionnaires,lehtinen2023automatedQuestionsHelp,santos2022jask}
like Java with questions such as~\cite{lehtinen2021askAutomatically}
``Line [N] uses a variable. Enter the line number where that variable
is declared.'' However, nowadays the first point of contact for many
learners is block-based
programming with environments such as
\Scratch~\cite{maloney2010,mcgill2020}, where questions like the above example 
are not feasible:
In \Scratch variables are created directly in the user
interface rather than via declaration statements, and blocks cannot be
referenced by line numbers.

In this paper, we introduce automatic generation of QLCs for \Scratch
programs. \Cref{fig:code} shows an excerpt of a \Scratch game that
repeatedly repositions a sprite until the player has caught it five
times. Although the code is correct, a student may have produced it
through trial and error or tinkering, they may have copied it from a
colleague, or they may have asked a large language model to generate
it. To check whether the concept of a conditional loop was not just
used but also understood, we generate the question ``\textit{Which of
  these blocks controls how many times this loop is executed?}'',
highlighting the target of the question (\cref{fig:code}) as well as a
choice of five different answers (\cref{fig:question}). All five
blocks are plausible as they are contained in the script, but an
understanding of conditional loops only allows one of the choices. A
student's answer thus provides evidence of whether or not they possess
such an understanding.

In detail, the contributions of this work are:
\begin{itemize}
\item We systematically derive 30 types of QLCs for \Scratch using an
  established code comprehension model.
\item We extend the open source tool \litterbox for generating these
  QLCs automatically.
\item We demonstrate the \emph{applicability} of the tool and our QLCs
  using a dataset of \numprojects publicly shared \Scratch programs.
\item We demonstrate the \emph{plausibility} of \Scratch QLCs by
  showing that the performance of students at solving a programming
  task correlates to their ability to answer QLCs.
\end{itemize}
Our investigation shows that the proposed QLCs can be generated
frequently in practice, and we find that the number of correctly
answered QLCs correlates to the performance at writing code. These
encouraging initial findings suggest that automatically generated QLCs
can be a practical means when using \Scratch to assess students’
program comprehension and to provide teachers with insights into the
learning progress of their students.



\section{Background}
\label{sec:background}

Current question generators are not applicable to \Scratch. We
therefore design new questions using an established code comprehension
model, and then implement static program analysis to generate
these questions automatically.

\begin{table}[t]
\centering
\caption{The Block Model~\cite{schulte2008blockModel}: columns 
represent different program dimensions, rows represent the scope of 
focus.}
\label{block_model}
\sffamily
\small
\begin{tabularx}{\columnwidth}{>{\color{white}\columncolor[gray]{.4}}lXXX}
\rowcolor[gray]{.4}  & \color{white}Text & \color{white} Execution & \color{white} Purpose \\
\rotatebox[origin=c]{90}{\begin{minipage}[c]{1cm}\centering{}Atom\end{minipage}} &    Language elements & Operation of an element & Purpose of an element \\
\rotatebox[origin=c]{90}{\begin{minipage}[c]{1cm}\centering{}Block\end{minipage}} & Synt. / sem\-an\-tically related regions & Operation of a block of 
code & Function of a block of code \\
\rotatebox[origin=c]{90}{\begin{minipage}[c]{1cm}\centering{}Relation\end{minipage}} & References between blocks of code & Flow between blocks of code& How 
goals are 
related to sub-goals \\
\rotatebox[origin=c]{90}{\begin{minipage}[c]{1cm}\centering{}Macro\end{minipage}} & Overall structure of the program text & Algorithm or program behaviour & Goal or purpose of the program \\
              
\end{tabularx}
\end{table}

\subsection{Questions about Learner's Code}

Questions about Learner’s Code (QLCs) are defined as questions
referring to concrete constructs or patterns in the student’s own
program~\cite{lehtinen2021askAutomatically}.  QLCs are designed to
determine code comprehension
capabilities~\cite{lehtinen2021askAutomatically,santos2022jask} by
referring to specific parts of a learner's code, which corresponds to
the second level in Bloom's taxonomy~\cite{anderson2001bloomRevised}
(\emph{Comprehension}).  As QLCs target program understanding rather
than skills related to writing programs they usually do not extend
into the \emph{Application} level.  Nevertheless, correctly answered
QLCs have been shown to correlate with success and
retention~\cite{lehtinen2021struggle}, thus indicating usefulness for
detecting weak prerequisite
knowledge~\cite{lehtinen2023automatedQuestionsHelp} and as a means to
assess programming skills. To support educators who want to use QLCs,
they can be generated
automatically~\cite{lehtinen2021askAutomatically}.



\subsection{The Block Model} 

 What constitutes program understanding has been captured in different
models~\cite{mayrhauser1995comprehension, pennington1987stimulus,
  soloway1984knowledge,burkhardt2002object,storey2005theories}. The
\emph{Block Model}~\cite{schulte2008blockModel} is one particular
model that other models can be directly mapped
to~\cite{schulte2010introduction}, and which was used in previous work
on automated generation of
QLCs~\cite{lehtinen2021askAutomatically,lehtinen2023automatedQuestionsHelp,lehtinen2021struggle,santos2022jask}.
\Cref{block_model} shows the block model, where columns represent
different program dimensions (the text, the execution, or the
purpose), while rows represent the scope of focus (individual
elements, blocks of related elements, relations between blocks of
elements, and the program as a whole). While existing QLCs may not be
applicable to \Scratch, the block model itself is nevertheless
suitable for developing new QLCs for \Scratch.

\subsection{Analysing \Scratch Projects}

Generating QLCs automatically requires analysing the code of a given
target program. \Scratch code consists of visual blocks that represent
programming commands in colours representing categories, and shapes
determining their type as well as how they can be assembled. For
example, stackable blocks represent statements and diamond-shaped or
round blocks represent expressions. To analyse \Scratch programs, they
first need to be converted to a more traditional intermediate
representation such as abstract syntax trees (AST), which captures the
different blocks and their relations. This is done by static analysis
tools like \litterbox~\cite{fraser2021litterbox}, enabling
applications such as finding code smells and bugs~\cite{fraedrich2020}
or code perfumes~\cite{obermueller2021perfumes} in \Scratch code,
which are helpful in assessing code rather than learners.
However, the same technical approach can serve as a means for
generating QLCs for \Scratch code automatically.


\section{Questions on \Scratch Code}
\label{sec:approach}

\begin{table*}
  \centering
  \caption{QLCs sorted into aspects of the Block Model.}
  \label{questions_block_model}
  \sffamily
  \small
  \begin{tabularx}{\textwidth}{>{\color{white}\columncolor[gray]{.4}}lXXX}
\rowcolor[gray]{.4}  & \color{white}Text & \color{white} Execution & \color{white} Purpose \\
		\rotatebox[origin=c]{90}{\begin{minipage}[c]{1cm}\centering{}Atom\end{minipage}} &     \makecell{\circled{1}\multipletype Block Controlling Loop \\
					\circled{2}\multipletype Element in Loop Condition \\
					\circled{3}\multipletype If Block Condition \\
					\circled{4}\stringtype Variable in Script}  & \circled{5}\stringtype Set Variable & 
					\makecell{\circled{6}\freetype Purpose of If Condition \\
					\circled{7}\freetype Purpose of Loop Condition\\
					\circled{8}\freetype Purpose of Variable}\\
		\rotatebox[origin=c]{90}{\begin{minipage}[c]{1cm}\centering{}Block\end{minipage}} & \makecell{\circled{9}\multipletype Blocks in If Statement \\
				\circled{10}\multipletype Element in Loop Body}  
				& \makecell{\circled{11}\multipletype If-Else Statement Execution\\
				\circled{12}\yesnotype If-Then Statement Execution \\
				\circled{13}\numbertype Repeat Times Execution}
				& 
		 \makecell{\circled{14}\freetype Purpose of Forever Loop \\
		\circled{15}\freetype Purpose of My Block \\
		\circled{16}\freetype Purpose of Repeat Times Loop \\
		\circled{17}\freetype Purpose of Repeat Until Loop \\
		\circled{18}\freetype Purpose of Script} \\
		\rotatebox[origin=c]{90}{\begin{minipage}[c]{1cm}\centering{}Relation\end{minipage}} &  \makecell{\circled{19}\multipletype My Block Definition \\
		\circled{20}\multipletype Script to Set Variable \\
		\circled{21}\multipletype Statement Triggers Event} & \makecell{\circled{22}\numbertype Scripts Triggered by Event \\
		\circled{23}\multipletype Scripts Triggered by Statement} &      \circled{24}\freetype Purpose of Broadcast \\
		\rotatebox[origin=c]{90}{\begin{minipage}[c]{1cm}\centering{}Macro\end{minipage}} & \makecell{\circled{25}\numbertype Scripts for Actor \\
		\circled{26}\numbertype Scripts in Program \\ \circled{27}\multipletype Variable for Actor }    
		& \makecell{\circled{28}\multipletype Script Execution Order Different Actors \\
		\circled{29}\yesnotype Script Execution Order Same Actor }   & \circled{30}\freetype Purpose of Program \\
		
	\end{tabularx}
\end{table*}

The approach of generating QLCs for \Scratch code automatically
consists of two distinct aspects: First, suitable question types need
to be designed, and second, an approach to derive instances of these
questions for a given \Scratch program needs to be developed.

\subsection{Designing QLCs for \Scratch}

To design questions that will aid students’ program comprehension we
use the Block Model as a guide, but therefore we have to adapt it to
the special features of a block-based language. Despite the
differences between text- and block-based languages, for each program
aspect represented in the Block Model (\cref{block_model}), we can
find a similar construct in \Scratch programs:
\begin{itemize}
  \item \emph{Atoms} relate to basic language elements, for which we consider
questions about variables, conditions, and basic value elements such
as literals and reporter blocks.
  \item \emph{Blocks} relate to questions about syntactically or semantically
related regions of code, which in \Scratch includes loops, conditional
statements, scripts, and procedures.
  \item \emph{Relations} deal with references between blocks of
code. Questions in this category cover concepts such as the definition
of a procedure, statements that change the value of a variable,
broadcast messages, and events which cause scripts to run.
\item \emph{Macros} look at the program as a whole. These questions
  consider larger concepts like the scripts or variables in a program,
  or the execution order of scripts.
\end{itemize}

Questions are furthermore sorted into five different question types
depending on the type of answer they require:
\begin{itemize}
  \item \emph{Number:} the answer is a single number; denoted by \numbertype.
  \item \emph{Strings:} the answer is one or more strings; denoted by 
\stringtype.
  \item \emph{Yes/No} the answer is either "yes" or "no"; denoted by 
\yesnotype.
  \item \emph{Multiple Choice:} the user is shown a set of choices and must 
choose
which ones are the correct answer. There may be more than one correct 
choice; denoted by \multipletype.
  \item \emph{Free Text:} the question is open-ended and requires the user to 
provide
some explanation as their answer; denoted by \freetype.
\end{itemize}

\subsection{QLCs for \Scratch}

The QLCs for \Scratch are summarised in \cref{questions_block_model}.
The individual questions are described in the following, sorted by
their scope in the Block Model starting in the Atom-Text field:

\subsubsection{Atom}

\parag{\circled{1}\multipletype Block Controlling Loop}: \textit{Which
  of these blocks controls how many times this loop is executed?} This
question checks if a user recognises a loop condition based on a
description of its function.  In
\inlineFigure{0.2}{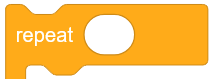} loops, the loop condition
is an oval block representing a number value, such as a number
literal. The condition of a
\inlineFigure{0.13}{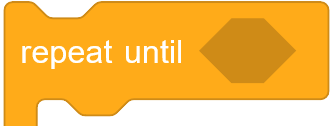} loop, on the other hand,
is a boolean block. So, the set of alternate choices presented to the
user include any non-control blocks, any boolean blocks, and any
number literals found in the script. The correct answer will be a
single boolean block or number literal, depending on which type of
loop is being referenced.

\parag{\circled{2}\multipletype Element in Loop Condition:}
\textit{Which of these elements are part of a loop condition?}  This
question is similar to \circled{1}, but focuses on the blocks inside
the condition and not the whole condition. It requires at least one
loop with condition, and the script must also contain at least one
non-boolean expression that is not in a loop condition in order to
show alternate choices.  In case there is more than one loop with a
condition, the correct answer includes elements present in all loop
conditions rather than one specifically.

\parag{\circled{3}\multipletype If Block Condition:} \textit{Which
  of these blocks is the condition of this if block?} This question
checks recognition of a condition in conditional statements. It
presents non-control blocks and boolean blocks as choices, including
the correct boolean block.

\parag{\circled{4}\stringtype Variable in Script:} \textit{Give
  the name(s) of the variable(s) in this script.} This question
assesses whether a student is able to recognise variables. One
question is created for every script with at least one variable. The
answer is a list of names of all variables present.

\parag{\circled{5}\stringtype Set Variable:} \textit{What value will
  (variable) have after this statement is executed?} This question
assesses whether the user
understands \begin{scratch}[scale=0.4]\blockvariable{set
    \selectmenu{variable} to \ovalnum{}}\end{scratch} blocks.  Since
variables can take numeric and text values, a string answer is
expected to accommodate both options.

\parag{\circled{6}\freetype Purpose of If Condition:} \textit{What is
  the purpose of the condition in this if statement?} This asks to
describe the role of a condition in an
\inlineFigure{0.13}{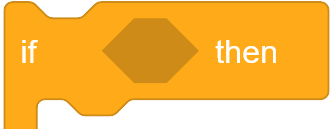} block. Free text answers are
open-ended, and teachers might expect different responses depending on
the curriculum.

\parag{\circled{7}\freetype Purpose of Loop Condition:} \textit{What
  is the purpose of the condition in this loop?} This question is
generated for \inlineFigure{0.13}{figures/blockRepeatUntil} blocks and
asks the purpose of the boolean condition in the loop.

\parag{\circled{8}\freetype Purpose of Variable:} \textit{What is
  the role of (variable) in this program?} This question asks
to describe the role of a given variable.

\subsubsection{Block}

\parag{\circled{9}\multipletype Blocks in If Statement:} \textit{Which
  of these blocks are found inside an if statement?} This question
checks recognition of the scope of an
\inlineFigure{0.13}{figures/blockIfThen} block by asking to identify
the statements inside it. A question is created for every script where
there is at least one \inlineFigure{0.13}{figures/blockIfThen} block
and at least one statement outside it. Since no specific
\inlineFigure{0.13}{figures/blockIfThen} block is referenced in the
question, statements inside any if statement in the script are
considered correct answers.

\parag{\circled{10}\multipletype Element in Loop Body:}
\textit{Which of these elements are found inside the body of a loop?}
This question checks if a user recognises a loop by asking them to
identify the elements inside of it. It differentiates between elements
in the loop condition and its body, as the former has its own question
(i.e., \textit{Element in Loop Condition}).  A question is created for
every script with at least one loop containing a non-boolean
expression, and at least one expression outside of a loop body. Since
no specific loop is referenced, elements inside any loop body in the
script are considered correct answers.

\parag{\circled{11}\multipletype If-Else Statement Execution:}
\textit{Which set of statements will be executed if <condition> is
  [true/false]?} This question checks understanding of the control
flow of an \textit{if else} statement and how it relates to the
condition. The user is shown two sets of blocks as choice
(\textit{then} and \textit{else} statements.) If either of these sets
is empty, the question is not generated. The question randomly assigns
true or false to the question text so that the expected answer can
vary.

\parag{\circled{12}\yesnotype If-Then Statement Execution:} \textit{In
  this if statement, will [statement] be executed if condition is
  [true/false]?} This question is similar to the previous one but for
\inlineFigure{0.13}{figures/blockIfThen} blocks with at least one
statement in their body, and one of these is selected randomly.

\parag{\circled{13}\numbertype Repeat Times Execution:} \textit{How
  many times is [statement] executed?} This question checks
understanding that statements within a loop are executed
repeatedly. The question considers
\inlineFigure{0.2}{figures/blockRepeatTimes} statements with a literal
value as the loop condition so that the answer can be deduced directly
from the loop statement. A question is generated once for every
\inlineFigure{0.2}{figures/blockRepeatTimes} block. However,
\inlineFigure{0.2}{figures/blockRepeatTimes} blocks nested inside
another loop are not considered, as this could lead to confusion about
how many times a statement is executed.

\parag{\circled{14}\freetype Purpose of Forever Loop:} \textit{Explain
  the purpose of this loop.} This question asks to explain the purpose
of a \inlineFigure{0.13}{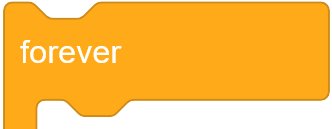} block.

\parag{\circled{15}\freetype Purpose of My Block:} \textit{Explain
  the function of this My Block.} This question asks the user to
explain a self defined procedure.

\parag{\circled{16}\freetype Purpose of Repeat Times Loop:}
\textit{Explain the purpose of this loop.} The user is asked to
explain the purpose of a \inlineFigure{0.2}{figures/blockRepeatTimes}
block.

\parag{\circled{17}\freetype Purpose of Repeat Until Loop:}
\textit{Explain the purpose of this loop.} Same as previous, but for
\inlineFigure{0.13}{figures/blockRepeatUntil} blocks.

\parag{\circled{18}\freetype Purpose of Script:} \textit{What is
  the purpose of this script?} This question asks the user to discuss
the purpose of a script.

\subsubsection{Relation}

\parag{\circled{19}\multipletype My Block Definition:}
\textit{Which of these shows the definition of this [My Block]?} This
question determines whether the user can identify the definition of a
procedure. They are shown a script containing a procedure call
statement, and given a set of scripts and procedure definitions to
choose from. The question is generated for each defined procedure
which is called at least once.

\parag{\circled{20}\multipletype Script to Set Variable:}
\textit{Which script sets the value of (variable)?} This question asks
the user to identify a script where the value of a variable is
set. Rather than having a question that simply asks the user to
identify a \begin{scratch}[scale=0.4]\blockvariable{set
    \selectmenu{variable} to \ovalnum{}}\end{scratch} block, this
question is designed to emphasise the relationship between a variable
and the script where its value is set, in an attempt to simulate a
question about variable initialization. Multiple scripts are shown,
the correct answer is the script containing the
fitting \begin{scratch}[scale=0.4]\blockvariable{set
    \selectmenu{variable} to \ovalnum{}}\end{scratch} block.

\parag{\circled{21}\multipletype Statement Triggers Event:}
\textit{Which of these statements will cause this script to start?}
One way in Scratch that program elements reference one another is
through broadcast messages. When a broadcast message is executed, any
scripts beginning with the corresponding event will run. Similarly, a
script beginning with the
event \begin{scratch}[scale=0.4]\blockinit{When backdrop switches to
    \selectmenu{my backdrop}}\end{scratch} will be triggered by the
statement \begin{scratch}[scale=0.4]\blocklook{switch backdrop to
    \ovallook*{my backdrop}}\end{scratch}. This question asks students to
identify the statements that will trigger a script to start. The
question requires there to be at least one triggering statement in
some script and another script which starts with the corresponding
event.

\parag{\circled{22}\numbertype Scripts Triggered by Event:}
\textit{How many scripts will run in the whole project when [event]
  happens?} This question is about the relationship between events and
script execution, and requires referring back to the full program.  A
relevant event (e.g., green flag click) is referenced in the question
text, and the user has to answer how many scripts will be
executed. Note that only the immediately triggered scripts are
considered here but no further propagation by events triggered in
these.  This also applies to the next QLC.

\parag{\circled{23}\multipletype Scripts Triggered by Statement:}
\textit{Which scripts will run after this [statement] is executed?} This
question is similar to the last but instead looks at scripts whose
events are triggered by statements in other scripts. As such, it is
the counterpart to \textit{Statement Triggers Event}. These statements
include \inlineFigure{0.1}{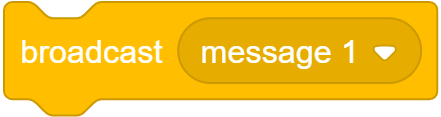},
\inlineFigure{0.1}{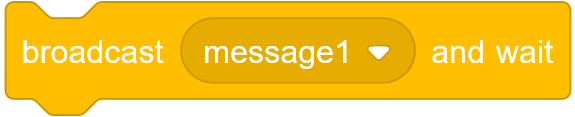},
\begin{scratch}[scale=0.4]\blocklook{switch backdrop to
		\ovallook*{backdrop}}\end{scratch}, and 
\begin{scratch}[scale=0.4]\blocklook{switch backdrop to
    \ovallook*{backdrop} and wait}\end{scratch}.

\parag{\circled{24}\freetype Purpose of Broadcast:} \textit{What
  happens when this broadcast message is sent?} Broadcast statements
are associated to a message. We generate a question for each message
that is found in a broadcast statement, rather than one for every
broadcast statement.

\subsubsection{Macro}
\parag{\circled{25}\numbertype Scripts For Actor:} \textit{How
  many scripts does [actor] have?} This question checks if the user
can identify the scripts belonging to a specific actor. It is
generated once for each actor.

\parag{\circled{26}\numbertype Scripts in Program:} \textit{How
  many scripts are in the whole program?} Similar to \circled{25};
generated once for the project.

\parag{\circled{27}\multipletype Variable For Actor:} \textit{Which
  sprite does (variable) belong to?} This assesses understanding of
local vs. global variables.

\parag{\circled{28}\multipletype Script Execution Order Different
  Actors:} \textit{The first script belongs to [actor1] and the second
  script belongs to [actor2].  Suppose [actor2] is in front of
  [actor1]. When [event] happens, which script will run first?} This
question requires an understanding of the execution order of
scripts. When some event happens, all scripts beginning with the
corresponding event block will be triggered. The execution order of
scripts belonging to different actors are dependent on the layer the
actor is on.  Sprites on a higher layer execute scripts first.

\parag{\circled{29}\yesnotype Script Execution Order Same Actor:}
\textit{These scripts belong to the same sprite. When [event] happens,
  is it possible to tell which script will run first?} This question
is similar to the last but instead looks at scripts belonging to the
same actor. When considering the scripts belonging to the same actor,
the execution order of these scripts is random. The answer to this
question is always ``No.''

\parag{\circled{30}\freetype Purpose of Program:} \textit{Describe
  what this program does.} This question asks the user to discuss the
program as a whole to demonstrate high level program comprehension.

\subsection{Automated Generation}


To automatically generate QLCs for \Scratch projects, we extended the
\litterbox tool~\cite{fraser2021litterbox}, which parses programs to
an abstract syntax tree (AST), and then provides a convenient
visitor-pattern for AST traversal. Each of our 30 questions is
implemented as one such \emph{question finder}, which traverses the
AST and returns all possible instances of that question type that can
be instantiated for the given program. For multiple choice questions
we also generate alternative answer options during traversal.
%
\litterbox returns a JSON-File containing questions as text snippets including
ScratchBlocks\footnote{\url{https://github.com/scratchblocks/scratchblocks},
	last accessed 17.03.2025}
syntax for code. Questions that target specific
parts of a script highlight these (see \cref{fig:code}), unless
the whole script has to be considered for answering the question. For
questions targeting a whole actor or project, no code is
shown. Questions will be generated each time the corresponding blocks
appear.

\subsection{Eliciting and Assessing Responses}

This paper presents the \litterbox extension which automatically
extracts questions from \Scratch programs. Once extracted, these
questions need to be presented to learners, and their responses need
to be assessed. While outside the scope of this paper, we implemented
a basic prototype extension of the \litterbox web user interface that
displays questions and the targeted script as seen in \cref{fig:code}
as a proof of concept. Ideally, however, we envision integration into
existing learning management systems for deploying the tool in the
classroom.
For some of the questions assessment can be automated: The responses
to Number, Strings, Multiple Choice, and Yes/No questions are all
amenable to automatic checking, since the correct answer among the
answer choices is provided by our \litterbox extension.  Free Text
questions, however, are open-ended and, therefore, the answers need to
be checked manually for now. 


\section{Evaluation}

In order to study the ability to generate useful QLCs about \Scratch code 
automatically, we investigate the following research questions:
\begin{itemize}
	\item \textbf{RQ1:} How frequently are different QLCs generated?
	\item \textbf{RQ2:} Does students' performance at answering questions 
	relate to their performance at coding?
\end{itemize}

\subsection{Experimental Procedure}
\subsubsection{RQ1:}
To evaluate applicability, we investigate whether each of the 30
question types occurs in practice, and how many such questions are
typically generated. We use a randomly sampled dataset of \numprojects
publicly shared \Scratch projects, excluding empty or remixed projects
to avoid skewing our analysis of question frequency.
For each QLC, we look at the cumulative total of instances generated
for the \numprojects projects, as well as the number of projects where
at least one instance of the question was found.

\subsubsection{RQ2:}
To evaluate plausibility, we conducted a classroom experiment with 34
ninth grade students from Germany, who already had prior experience
with \Scratch. The students received a scaffolding \Scratch project
for the Boat Race
game,\footnote{\url{https://projects.raspberrypi.org/en/projects/boat-race},
  last accessed 17.03.2025} including the boat movement as well as the
winning and loosing functionalities. Before starting they had to
answer ten QLCs about the scaffolding code, which we provided on paper
to avoid confusing them with an unfamiliar user interface and adding
cognitive load. We used all questions generated for the scaffolding
code except for those requiring free text responses or combined blocks
to simplify evaluation.
After answering the QLCs the students were tasked to add seven new
functionalities until the end of the lesson.

We hypothesise that students with better program comprehension skills
will perform better at implementing these
functionalities. Consequently, we would expect students who are better
at answering QLCs about the scaffolding code to produce better
solutions.
To quantify the performance at answering questions, points are
assigned to each question depending on their answer.  Since the first
two questions are multiple choice, for these questions we assign $0.2$
points for each correct response as there were 5 possible choices for
each question. Note that not ticking a wrong choice also is considered
a correct response.  Questions 3--10 have an answer that is either
right or wrong; these questions are given 1 point for a correct
answer.  The total points for each student are added up, and then
divided by 10 to achieve their final score.
To quantify the performance at the coding task, we use
\whisker~\cite{stahlbauer2019testing}, an automated testing framework
for \Scratch programs.  We created seven tests for the extended Boat
Race to cover all functionalities. Using these tests, each submission
is assigned a score of $x/7$ where $x$ is the number of passed tests.
Finally, the two scores are compared calculating the Pearson
correlation coefficient with $\alpha=0.05$.


\subsection{Threats to Validity}

Threats to \textit{internal validity} may arise from the available
time being too short for some participants; however, we had to fit the
experiment into one lesson and prior usage of the coding task in other
scenarios suggests that one lesson is sufficient. Threats to internal
validity may also result from bugs in our modifications to \litterbox
by adding the question generation; however, we thoroughly tested the
setup. Besides the number of students participating, threats to
\textit{external validity} may also result since we used one task in
the classroom experiment and only publicly shared projects in the
mined data set, limiting generalisation of results.  Threats to
\textit{construct validity} may result since we created QLCs
beforehand in the classroom experiment, rather than creating them on
the fly. The students’ parents were informed about the programming
course and all respondents consented to anonymised data use.

\begin{table}
\centering
\caption{Number of questions instances found in total and on average, and 
number of projects showing each question.}
\vspace{-1em}
\label{instances_projects}
\begin{tabular}{@{}lrrr@{}}
\toprule
                          Question Name & \multicolumn{2}{r}{Questions}
                           &       \# Projects \\
                          & Total \# & $\varnothing$&\\
\midrule
                 Block Controlling Loop &   1,271,938 &                     5.0 &  252,443 (42.0\%) \\
                 Blocks in If Statement &     815,051 &                     4.3 &  188,448 (31.4\%) \\
                   Element in Loop Body &   1,959,636 &                     5.8 &  337,068 (56.1\%) \\
              Element in Loop Condition &   1,203,119 &                     5.0 &  242,806 (40.4\%) \\
                     If Block Condition &   1,796,277 &                     7.6 &  237,505 (39.5\%) \\
            If-Else Statement Execution &     817,975 &                     7.7 &  105,581 (17.6\%) \\
            If-Then Statement Execution &   2,789,980 &                    12.8 &  217,799 (36.2\%) \\
                    My Block Definition &     368,712 &                     5.4 &   67,796 (11.3\%) \\
                   Purpose of Broadcast &   1,103,245 &                     5.9 &  187,198 (31.2\%) \\
                Purpose of Forever Loop &   2,311,690 &                     6.5 &  355,013 (59.1\%) \\
                Purpose of If Condition &   5,103,322 &                    20.7 &  246,997 (41.1\%) \\
              Purpose of Loop Condition &     486,902 &                     5.3 &   91,368 (15.2\%) \\
                    Purpose of My Block &     530,973 &                     7.4 &   71,885 (12.0\%) \\
                     Purpose of Program &     600,913 &                     1.0 & 600,913 (100.0\%) \\
           Purpose of Repeat Times L. &   2,023,993 &                     8.7 &  232,665 (38.7\%) \\
           Purpose of Repeat Until L. &     486,902 &                     5.3 &   91,368 (15.2\%) \\
                      Purpose of Script &   9,748,844 &                    16.2 & 600,913 (100.0\%) \\
                    Purpose of Variable &   1,103,801 &                     5.2 &  212,969 (35.4\%) \\
                 Repeat Times Execution &   1,510,406 &                     7.8 &  193,543 (32.2\%) \\
Script Exec. Order Diff. Act. &     968,422 &                     2.5 &  
389,217 (64.8\%) \\
      Script Exec. Order Same Act. &     850,247 &                     3.3 &  
      255,330 (42.5\%) \\
                 Script to Set Variable &     594,269 &                     3.3 &  177,478 (29.5\%) \\
             Scripts Triggered by Event &   2,436,371 &                     4.1 &  594,798 (99.0\%) \\
         Scripts Triggered by Stmt.     &   1,071,663 &                     5.5 &  193,455 (32.2\%) \\
                      Scripts for Actor &   3,524,268 &                     5.9 & 600,913 (100.0\%) \\
                     Scripts in Program &     600,913 &                     1.0 & 600,913 (100.0\%) \\
                           Set Variable &   2,199,143 &                    12.1 &  182,477 (30.4\%) \\
               Statement Triggers Event &   1,110,725 &                     5.7 &  195,137 (32.5\%) \\
                     Variable for Actor &   2,404,752 &                     4.3 &  560,649 (93.3\%) \\
                     Variable in Script &   2,324,242 &                    10.9 &  212,969 (35.4\%) \\
\bottomrule
\end{tabular}
\vspace{-0.5em}
\end{table}

\subsection{RQ1: How Frequently Are Different QLCs Generated?}
\begin{table}
	\centering
	\caption{Percentage of projects with at least one question
          generated per Block Model cell.}
        \vspace{-1em}
	\label{block_model_projects}
	\sffamily
	\small
	\begin{tabularx}{\columnwidth}{>{\color{white}\columncolor[gray]{.4}}lrrr}
		\rowcolor[gray]{.4}  & \color{white}Text & \color{white} Execution & 
		\color{white} Purpose \\
		\begin{minipage}[c]{1cm}Atom\end{minipage}
                &    \cellcolor{cyan!64.5}\textsf{\atomTextProject} & \cellcolor{cyan!30.4}\atomExecutionProject & \cellcolor{cyan!49.0}\atomPurposeProject \\
                
		\begin{minipage}[c]{1cm}Block\end{minipage}
                & \cellcolor{cyan!61.2}\blockTextProject & \cellcolor{cyan!58.4}\blockExecutionProject & \cellcolor{cyan!100}\blockPurposeProject \\
                
		\begin{minipage}[c]{1cm}Relation\end{minipage}
		 & \cellcolor{cyan!46.5}\relationTextProject & \cellcolor{cyan!99}\relationExecutionProject & 
                 \cellcolor{cyan!31.2}\relationPurposeProject \\
                 
		\begin{minipage}[c]{1cm}Macro\end{minipage}
		 & \cellcolor{cyan!100}\macroTextProject & \cellcolor{cyan!71.7}\macroExecutionProject & \cellcolor{cyan!100}\macroPurposeProject \\
		
	\end{tabularx}
	\vspace{-1.5em}
\end{table}

\Cref{instances_projects} shows, for each QLC, the total number of question 
instances, as well as the number of projects that contain at least 
one instance of the question. Each of the \numprojects projects generated at 
least five questions, and there were a total of \numquestions generated 
questions.

Considering the totals, the most commonly generated questions are
\textit{Purpose of Script} (9,748,844), \textit{Purpose of If
  Condition} (5,103,322) and \textit{Scripts for Actor}
(3,524,268). These question types have no further constraints other
than existence of the relevant entities; all programs must have actors
and scripts, and \emph{if}-conditions appear to be frequently used.
In contrast, \textit{My Block Definition} (368,712), \textit{Purpose
  of Loop Condition} (486,902) and \textit{Purpose of Repeat Until
  Loop} (486,902) appear to relate to less frequent constructs.

\textit{Purpose of Program}, \textit{Purpose of Script},
\textit{Scripts for Actor} and \textit{Scripts in Program} rank first
in terms of the percentage of projects for which questions were
generated. The only requirement for these QLCs is the existence of at
least one script, which is true for all programs in our dataset.
%
%
\Cref{block_model_projects} illustrates the proportion of projects
covered by each cell of the Block Model; i.e., for how many of the
projects our approach generates at least one question. For
\emph{Block}--\emph{Purpose}, \emph{Macro}--\emph{Text}, and
\emph{Macro}--\emph{Purpose} questions were generated for \emph{all}
projects. For the other cells the questions depend on specific
programming constructs used, and we note a tendency for lower coverage
for lower level scopes of focus. This indicates that more questions at
a lower scope targeting \Scratch specific constructs like setting of
costumes in the \emph{Atom}--\emph{Execution} cell or the purpose of
clones in \emph{Purpose}--\emph{Relation} are needed to ensure
projects are covered well.
Furthermore, adding a \emph{None of the above} option could help increase the 
number of generated questions of the type Multiple Choice, as the requirements 
for selecting alternative answers could be lowered.

\summary{RQ1}{All questions can be generated frequently and at least
  top level questions can be generated for each project.}

\subsection{RQ2: Do Correct Answers Relate to Coding Performance?}

\begin{figure}[tb]
	\centering
	\includegraphics[width=0.5\columnwidth]{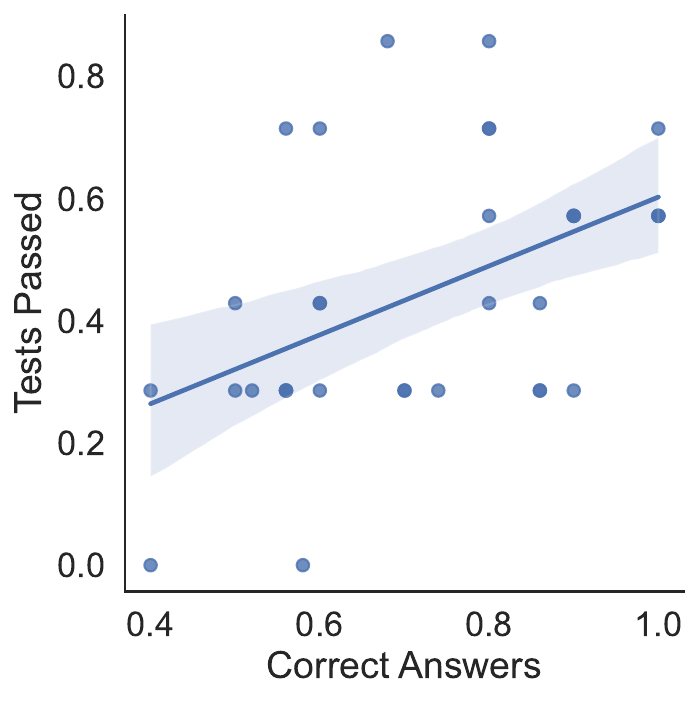}
        \vspace{-1em}
	\caption{\label{fig:rq2_scatter} Correlation between \whisker tests passed and 
	correctly answered QLCs ($r=0.467$, $p=0.005$).}
      \vspace{-1em}
\end{figure}

The relation between QLCs answered correctly prior to coding and the
number of functionalities correctly implemented are shown in
\cref{fig:rq2_scatter}. The Pearson correlation coefficient between
correctly answered questions and passed tests is $0.467$ ($p=0.005$),
i.e., a medium correlation. Assuming that program comprehension is
directly related to the ability to write
code~\cite{lister2010naturally,lister2009evidence,venables2009closerlook,lopez2008relationship},
this indicates that QLCs are a valid proxy for measuring program
comprehension.
The correlation will be influenced by the fact that no students
managed to implement all of the additional features, but this will not
only be influenced by program comprehension and coding skills, but
also by the limited time. Nevertheless, QLCs are a promising means to
detect missing prerequisite knowledge and
skills~\cite{lehtinen2023automatedQuestionsHelp}.

\summary{RQ2}{The performance at answering questions before completing
  programming tasks correlates to actual performance in successfully
  completing coding tasks.}


\section{Conclusions}

In order to assess whether students understand the \Scratch code they
wrote,
%
we introduced QLCs for \Scratch and the \litterbox tool to
automatically generate them. Our initial evaluation demonstrates that
these questions are frequently applicable; correlation between the
ability to answer these questions and writing code indicates that the
QLCs are a valid instrument for assessing program comprehension.
An important next step will be to evaluate the effects these QLCs on
learning and assessment, which will require an appropriate user
interface in which teachers can generate and select
questions. Although our 30 QLCs provide reasonable coverage of the
Block Model, there is potential for further questions, targeting
\Scratch specific constructs on lower scopes. While answers to most
question types can be checked automatically, this is challenging for
free text answers; however, we anticipate that large language models
provide an opportunity to automate this aspect.
To support the adoption of QLCs in \Scratch and further research,
\litterbox is available as open source at:
\textit{\url{https://github.com/se2p/LitterBox}}

\begin{acks}
  We thank Emily Courtney for her help in extending \litterbox and
  Ewald Wasmeier for his help conducting the classroom
  experiment. This work is supported by the DFG under grant \mbox{FR
    2955/5-1}.
\end{acks}

\balance

\bibliographystyle{ACM-Reference-Format}
\bibliography{library}

\end{document}